\documentclass{PoS}

\usepackage{epsf}
\usepackage{amscd}
\usepackage{amsmath}
\usepackage{graphicx,comment}
\input xy
\xyoption{all}

\def\ol#1{{\overline{#1}}}

\newcommand{\beq}{\begin{equation}}
\newcommand{\eeq}{\end{equation}}
\newcommand{\bea}{\begin{eqnarray}}
\newcommand{\eea}{\end{eqnarray}}

\newcommand{\nn}{\nonumber}
\newcommand{\benn}{\begin{displaymath}}
\newcommand{\eenn}{\end{displaymath}}

\def\mc#1{{\mathcal #1}}

\newcommand{\e}{\mathbf e}

\def\slashchar#1{\ensuremath{                               %
   \setbox0=\hbox{${}#1{}$}       
   \dimen0=\wd0                                 
   \setbox1=\hbox{/} \dimen1=\wd1               
   \ifdim\dimen0>\dimen1                        
      \rlap{\hbox to \dimen0{\hfil/\hfil}}      
      {}#1{}                                    
   \else                                        
      \rlap{\hbox to \dimen1{\hfil${}#1{}$\hfil}}   
      /                                         
   \fi}}                                        %

\title{Search for Chiral Fermion Actions on Non-Orthogonal Lattices }

\ShortTitle{Search for Chiral Fermion Actions on Non-Orthogonal Lattices }

\author{\speaker{Michael I. Buchoff}%
        \thanks{In collaboration with Paulo Bedaque, Brian Tiburzi, and Andr\'e Walker-Loud.}
        \thanks{This work is supported in part by the 
U.S.~
DOE
Grant No.~DE-FG02-93ER-40762.}\\
Maryland Center for Fundamental Physics, 
Department of Physics, 
University of Maryland, 
College Park,  
MD 20742-4111, 
USA \\
       E-mail: \email{mbuchoff@umd.edu}}

\abstract{The graphene-inspired fermion actions recently proposed by Creutz and Bori\c{c}i have sparked interest in the use of non-orthogonal lattices in lattice QCD. These fermion actions have the desired chiral symmetry and have the minimal doubling required by the Nielsen-Ninomiya no-go theorem. However, due to the lack of discrete symmetries, radiative corrections in the gauged lattice theory will lead to the generation of unwanted relevant and marginal operators. Other similarly motivated non-orthogonal fermion actions avoid these unwanted operators, but introduce incorrect continuum behavior or excessive fermion doubling.  A delicate balance of  symmetry is required for chiral symmetry, minimal doubling, and no relevant operators, and to date, no non-orthogonal lattice action has accomplished this balance.    }

\FullConference{The XXVI International Symposium on Lattice Field Theory \\
July 14 - 19, 2008\\
Williamsburg, Virginia, USA}

\begin{document}

\section{Introduction}
One major obstacle in various regularizations of lattice QCD is fermion doubling, where the continuum limit describes multiple fermions despite only one fermion being attached to each node.  While more complicated discretizations can reduce the amount of doubling, this often comes at the expense of exact chiral invariance as shown in the Nielsen-Ninomiya ``no-go'' theorem~\cite{nielsen}.   In the past, attempts were made by Karsten~\cite{karsten} and Wilczek~\cite{wilczek} to minimize the doubling (only two fermions) allowed by the Nielsen-Ninomiya ``no-go'' theorem while preserving chiral symmetry.  However, these actions broke additional symmetries leading to the generation of non-physical dimension three and four operators in the continuum limit.

Within the past year, Creutz~\cite{creutz} and, shortly after, Bori\c{c}i~\cite{borici} introduced a non-orthogonal graphene-inspired action in four-dimensions in order to achieve both minimal doubling and chiral symmetry.  Unfortunately, as shown in Ref.~\cite{bedaque}, this new action suffers from similar symmetry breaking issues that plagued the actions proposed by Karsten and Wilczek.   However, an intriguing realization upon studying the Bori\c{c}i-Creutz action is that if a four-dimension action contains the minimum $\mathbb{Z}_5$ permutation symmetry, the generation of the divergent dimension 3 operators could be prevented in the continuum limit.  As a result, Ref.~\cite{bedaque2}, explores several non-orthogonal lattices in order to achieve this symmetry, but these lattices also lead to other undesireable features.  These actions will be discussed throughout this note along with the Bori\c{c}i-Creutz action.

\section{Graphene}
The purpose of this section is to describe some of the most basic results of graphene lattices that have attracted a great deal of attention throughout the condenced matter community.  Graphene lattices, which are two dimensional hexagonal honeycomb lattices, have the intriguing property that massless fermions on the sites lead to exact Dirac fermions.

\begin{figure}[ht]
\center
\begin{tabular}{cc}
\includegraphics[width=0.5\columnwidth]{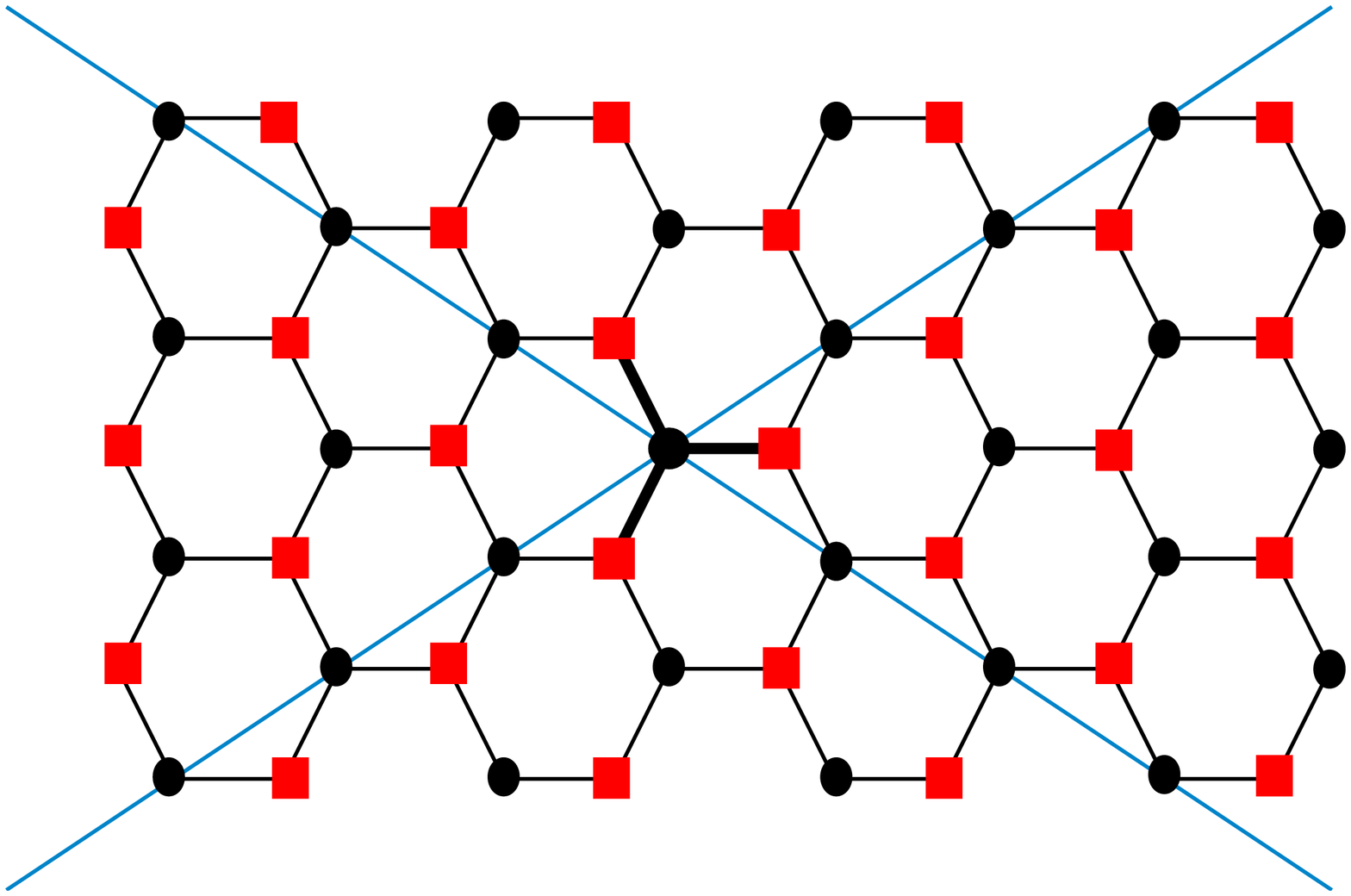}
\includegraphics[width=0.5\columnwidth]{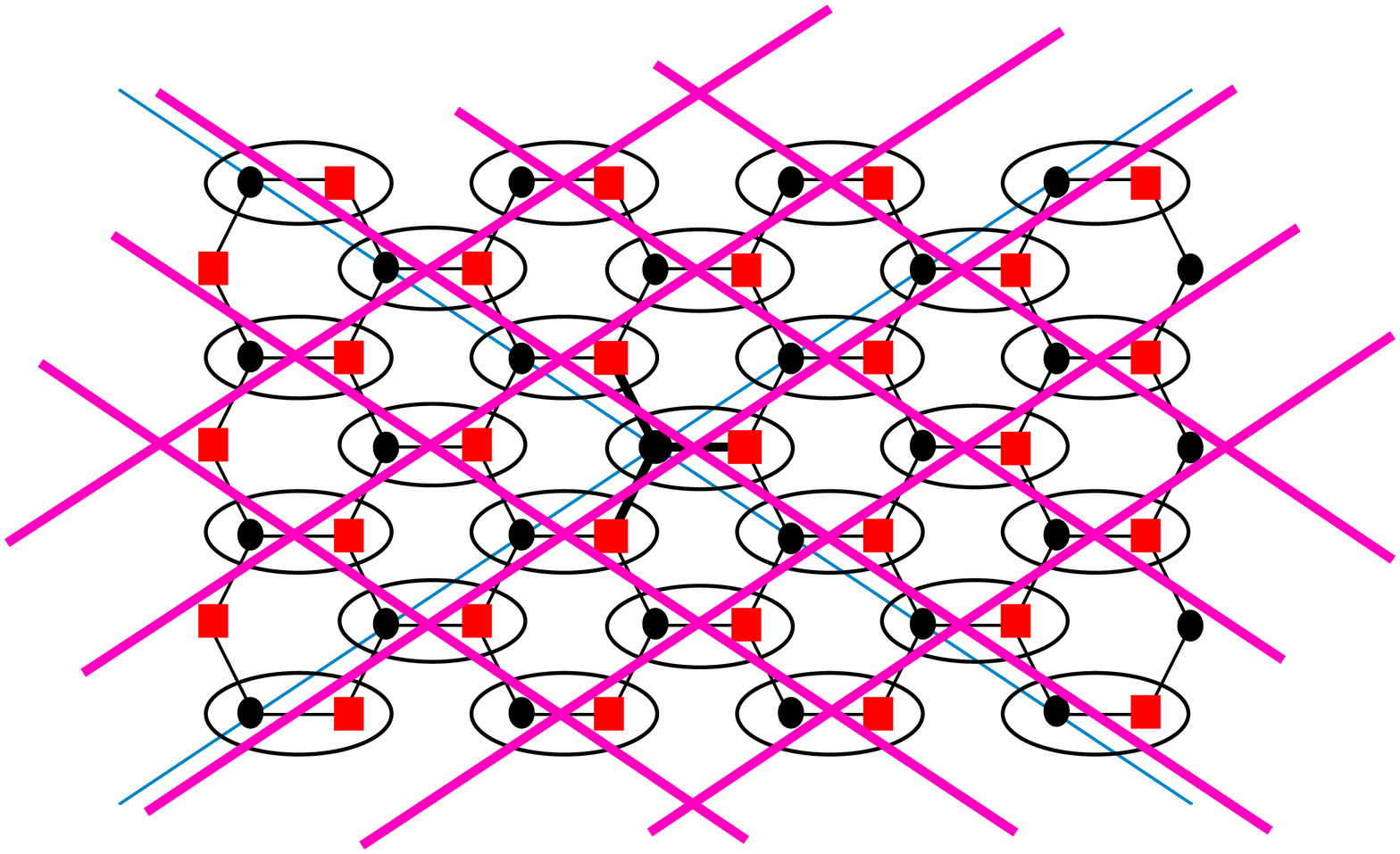}
\end{tabular}
\caption{(left) Graphene lattice composed of two sub-lattices.  The ``a-site'' lattice is represented by black circles and the ``b-site'' lattice is represented by red squares. (right) Illustrates the proposal by Creutz \cite{creutz} to group horizontal neighbors into two-atom sites, given by the ellipses.  With these sites, a grid can be constructed, which is given by the pink lines.}
\label{Grid}
\end{figure}
These lattices are analyzed by a two sub-lattice description.  As shown in Fig.~\ref{Grid}, one sub-lattice (often referred to as the ``a-sites'' or ``l-sites'') only communicates with its three nearest neighbors which are members of the other sub-lattice (often referred to as the ``b-sites'' or ``r-sites'').  Another important feature is that these three nearest neighbors possess a $\mathbb{Z}_3$ permutation symmetry, which means that these lattices are invariant under $120^\circ$ rotations.  As a result, the addition of the three primary vectors, $\e^\alpha$, created by the three connections to the neareast neighbors yields zero ($\sum_\alpha \e^\alpha = 0$).  Without going into detail, these features of a $1+1$ dimensional graphene lattice, along with two component Dirac spinors,  result in chiral Dirac fermions in the massless limit.  The question that is of interest to this note is how to extend this construction to four-dimensions.

\section{Graphene-Inspired Lattice Actions}

Within the past year, Creutz in Ref.~\cite{creutz} proposed a viable method of extending this construction to four dimensional lattice gauge theory.  The main idea behind this method was to combine the horizontally connected ``a'' and ``b'' sites into an individual two-atom ``site'' (as shown in Fig.~\ref{Grid} as the ellipses around these two-atom sites).  Through this process, the connecting segments between these two atom sides form a non-orthogonal grid, which is illustrated by the pink lines imposed on the lattice in Fig.~\ref{Grid}.  With this formulation,  Creutz generalized this non-orthogonal grid to four-dimensions while maintaining a construction similar to graphene.  

 \section{Bori\c{c}i-Creutz Action}
The action of this graphene-inspired, non-orthogonal grid in four dimensions proposed by Creutz \cite{creutz}, and later Bori\c{c}i \cite{borici} is given by 
\begin{eqnarray} 
S_{BC} 
&=& 
\frac{1}{2}\sum_{x}
\Bigg[
\sum_{\mu=1}^4
\left(\ol \psi_{x- \mu}
\, \e^\mu \cdot \Gamma \,
\psi_{x}
-
\ol \psi_{x+\mu}
\,
\e^\mu \cdot \Gamma^\dagger
\,
\psi_{x}
\right)
+
\ol \psi_x
\,
\e^5 \cdot \Gamma
\,
\psi_x-\ol \psi_x
\,
\e^5 \cdot \Gamma^\dagger
\,
\psi_x
\Bigg],
\label{eq:CreutzZ5}
\end{eqnarray}
where 
$\Gamma_\mu = (\vec{\gamma}, i \gamma_4)$ and the four-vectors $\e^\alpha$ are defined
in terms of two parameters $B$ and $C$ as
\bea\label{eq:es2}
\e^1 &=& \phantom{-}(\phantom{+}1,\phantom{+}1,\phantom{+}1,\phantom{4}B\ ),\nn\quad 
\e^2 = \phantom{-}(\phantom{+}1,-1,-1,\phantom{4}B\ ),\nn\quad
\e^3 = \phantom{-}(-1,-1,\phantom{+}1,\phantom{4}B\ ),\nn\\
\e^4 &=& \phantom{-}(-1,\phantom{+}1,-1,\phantom{4}B\ ),\nn\quad
\e^5 = -(\phantom{+}0,\phantom{+}0,\phantom{+}0,4BC).
\eea 
Note, each $\e^\alpha$ above is a four vector and the $\alpha$ does not refer to the individual components of this vector.  These $\e^\alpha$ vectors are generated from this non-orthogonal grid\footnote{See Ref.~\cite{bedaque2} for more details.}, which are related to the pink lines in Fig.~\ref{Grid}.  While written in a way that is similar to the na\"ive fermion lattice action, there are several key differences in this action.  First, the definition of $\Gamma_\mu$ contains an $i \gamma_4$, as opposed to the na\"ive fermion action that does not contain this factor of $i$.  This leads to an important sign difference in the fourth component of these vectors.  Second, these $\e^\alpha$ vectors contain the non-orthogonal structure of this lattice action, which lead to non-trivial linear combinations of Dirac matricies.    To understand the behavior of this lattice action, it is beneficial to write this action in momentum space, which is given by
\begin{equation}
\label{BC_Mom }
S_{BC}=
\int_p 
\ol\psi_p \left[ 
	i\sum_{\mu=1}^4 \bigg(\sin(p_\mu) \vec{\e}^\mu \cdot \vec{\gamma}
	+B\gamma_4(\cos(p_\mu)-C)\bigg)
\right] \psi_p.
\end{equation}
When written in momentum space, one can immediately confirm that this action has exact chiral symmetry.  In addition, for $B \neq 0$ and $0<C<1$, this action has two poles at 
\begin{eqnarray}
p_\mu^{(1)} &=& \tilde{p}(\phantom{+}1,\phantom{+}1,\phantom{+}1,\phantom{4}1\ )\nn\quad\quad
p_\mu^{(2)} = -\tilde{p}(\phantom{+}1,\phantom{+}1,\phantom{+}1,\phantom{4}1\ ),
\end{eqnarray}
where $\cos(\tilde{p}) = C$.  The existence of only two poles results in the action having minimal fermion doubling under the Nielsen-Ninomiya ``no-go'' theorem~\cite{nielsen}.  

While the features of exact chiral symmetry and minimal doubling are both convenient and attractive features in a lattice action, there are undesirable consequences as well.   These consequences stem from the action breaking additional discrete symmetries, namely parity ($\psi(\vec{p},p_4) \rightarrow \gamma_4 \psi(-\vec{p},p_4)$), charge conjugation ($\psi(\vec{p},p_4) \rightarrow C \bar{\psi}^T(\vec{p},p_4)$), and time reversal ($\psi(\vec{p},p_4) \rightarrow \gamma_5\gamma_4 \psi(\vec{p},-p_4)$).  These broken symmetries will lead to new unwanted operators that are a result of radiative corrections.

\subsection{Radiative Corrections}
The presence of additional broken symmetries in lattice actions allow for new operators to be generated through radiative corrections.  The new unphysical operators due to this lattice discretization fall into three categories; relevant, marginal, or irrelevant operators.  Irrelevant operators, which are operators of mass dimension five or higher, are proportional to positive powers of the lattice spacing $a$ and would vanish in the continuum limit, $a\rightarrow 0$.  The other two categories, relevant and marginal, remain in the continuum limit and add unphysical terms to the final results that require fine tuning to eliminate.  Marginal operators, dimension four, can have logarithmic lattice spacing dependence. Relevant operators, dimension three or less, are particularly bad since they are proportional to negative powers of $a$ and diverge in the continuum limit.  

The exact chirality of the Bori\c{c}i-Creutz action prevents the relevant operator from the Wilson action proportional to $a^{-1}\ol \psi \psi$, but the breaking of parity and time reversal allow to two additional relevant operators, $a^{-1}\sum_j c_3^{(j)} \mc O_3^{(j)}$, where
\begin{eqnarray}
\mc O_3^{(1)} &=& 4iB\ol \psi \gamma_4 \psi =i\sum_{\mu=1}^4 \ol \psi (\e^\mu \cdot \gamma) \psi, \nn\quad\quad
\mc O_3^{(2)} = 4iB\ol \psi \gamma_4 \gamma_5\psi = i\sum_{\mu=1}^4 \ol \psi (\e^\mu \cdot \gamma)\gamma_5 \psi.
\end{eqnarray}
An emphasis should be placed on the fact that $\mu = 1,2,3,4$ and does not include $e^5$ in this sum.   This illustrates the $S_4$ permutation symmetry the Bori\c{c}i-Creutz action possesses, which is apparent in this form since these relevant operators are invariant under any exchange of the four $\mu$ values.  In addition to these two relevant operators, there are additionally ten marginal operators\footnote{See Ref.~\cite{bedaque} for more detail on the marginal operators.}, which we will not focus on in this note.  All these operators will be generated unless the action possesses an additional symmetry. 

\subsection{Additional Symmetry?}
In physical two-dimensional graphene, there exists an additional $\mathbb{Z}_3$ rotational symmetry.  The question that now remains is whether an analogous symmetry exists for the graphene-inspired Bori\c{c}i-Creutz action.  Analysis on the subject \cite{bedaque2} shows that if an action in four dimensions  
poscesses the minimal $\mathbb{Z}_5$ permutation symmetry (or the larger two-operation $A_5$ or $S_5$ permutation symmetry), then for $\alpha = 1,2,3,4,5$, 
\begin{equation}
\label{sum_ea}
\sum_{\alpha=1}^5 \e^{\alpha} = 0,
\end{equation}
which results in the relevant operators   
\begin{eqnarray}
\mc O_3^{(1)} &=&i\sum_{\alpha=1}^5 \ol \psi (\e^\alpha \cdot \gamma) \psi = 0, \nn\quad\quad
\mc O_3^{(2)} =i\sum_{\alpha=1}^5 \ol \psi (\e^\alpha \cdot \gamma) \gamma_5\psi = 0.
\end{eqnarray}
Thus, a lattice action with this minimal symmetry would not have any relevant operators.  

\begin{figure}[ht]
\center
\begin{tabular}{cc}
\includegraphics[width=0.41\columnwidth]{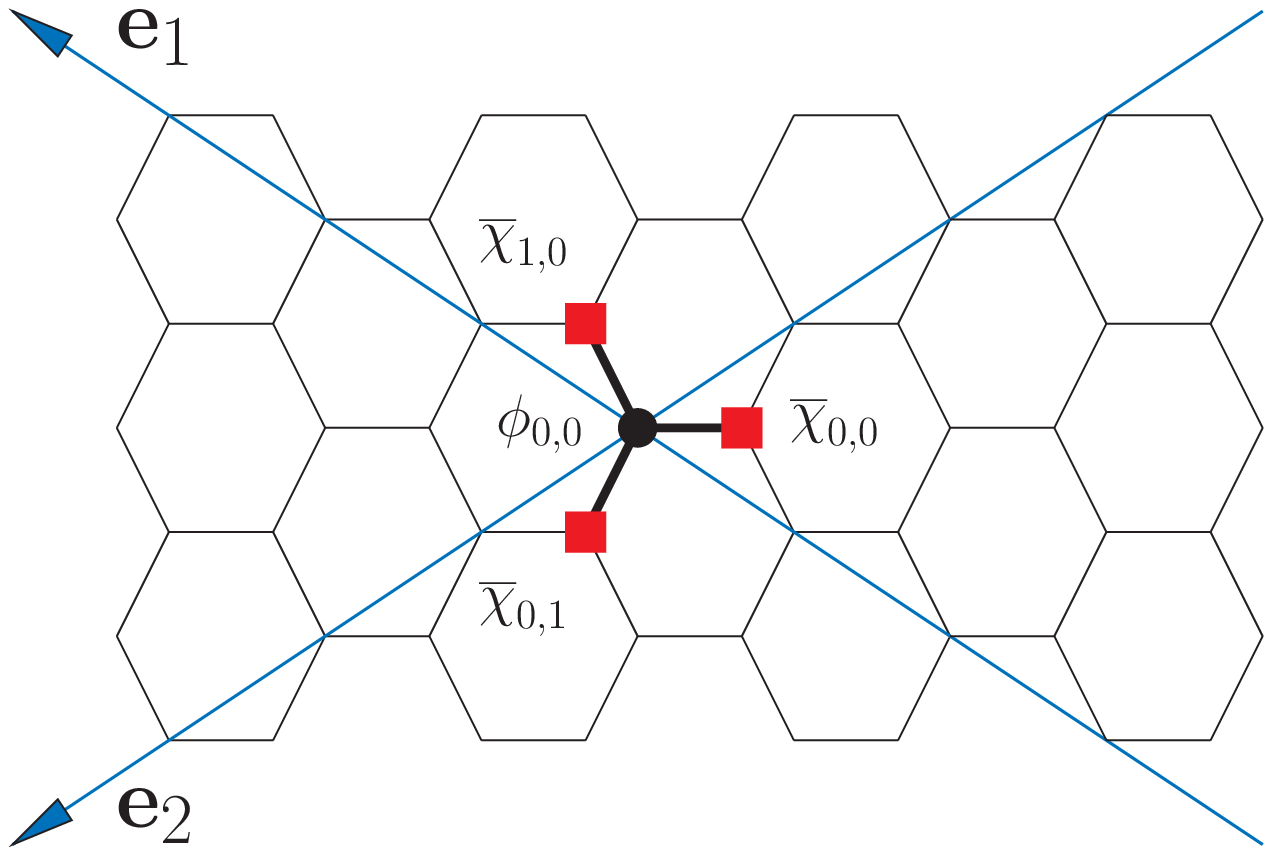} &
\includegraphics[width=0.41\columnwidth]{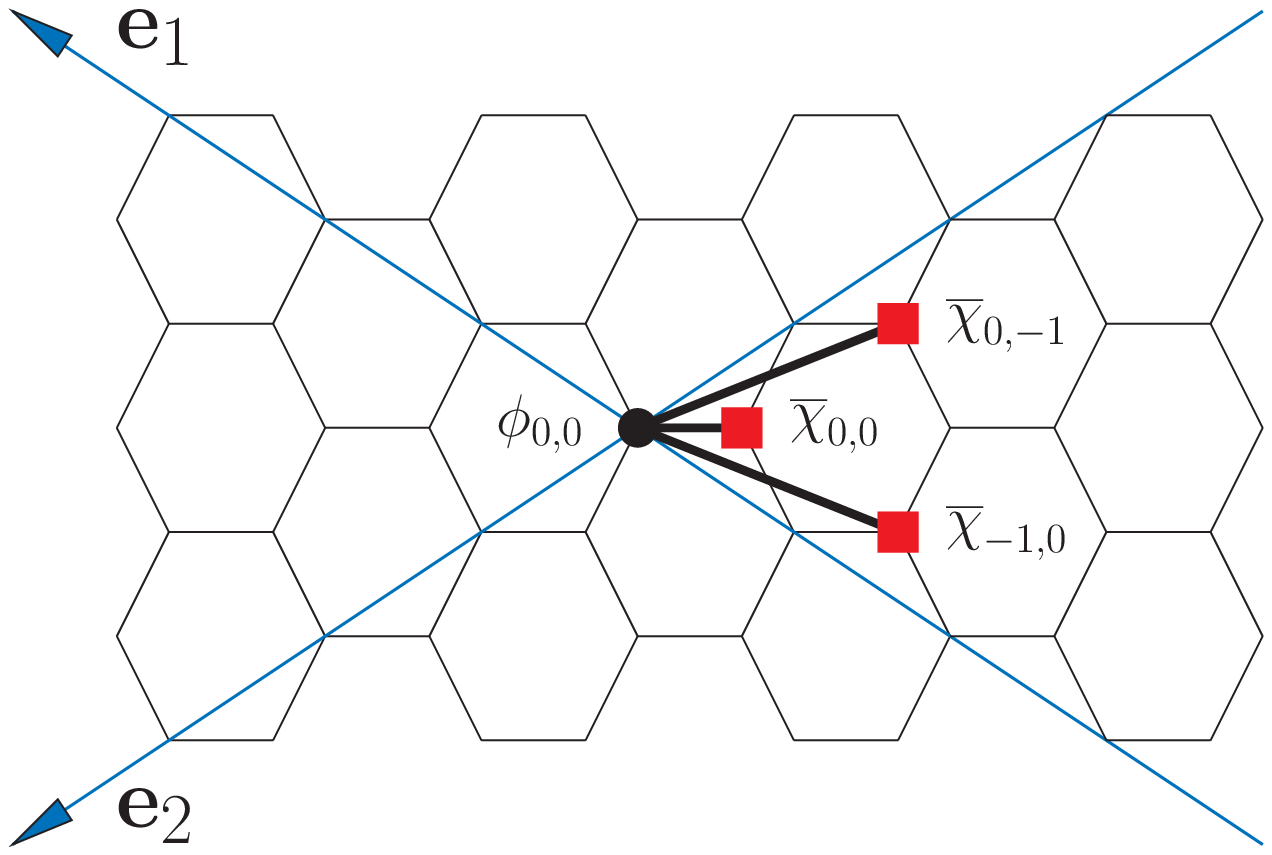}
\end{tabular}
\caption{Two-dimensional projection of  Bori\c{c}i-Creutz action with the blue duel basis vectors $\e_\mu = \e^\mu-\e^5$. (left) Projection of the first line in of the two-component action shows nearest neighbor interactions leading to $\mathbb{Z}_3$ rotational symmetry. (right) Projection of the second line shows next-to-nearest neighbor interactions breaking $\mathbb{Z}_3$.} 
\label{GB}
\end{figure}

So does the Bori\c{c}i-Creutz action have this minimal $\mathbb{Z}_5$ permutation symmetry?  To explore this question, it is useful to write the action in terms of two two-component spinors, one which acts as the ``a-site'' and one which acts as the ``b-site'' as illustrated in Fig.~\ref{Grid}.  The action written in this way is 
\bea\label{eq:BC_twocomponent}
S_{BC} 
&=& 
\frac{1}{2}\sum\limits_x 
\Bigg[
\sum\limits_{\mu=1}^4\left(
\ol\phi_{x-\mu} \, \Sigma \cdot \e^\mu \, \chi_{x}
 -\ol\chi_{x+\mu} \, \Sigma \cdot \e^\mu \, \phi_x  \right)
 +\ol\phi_{x} \, \Sigma \cdot \e^5 \, \chi_{x} 
 -\ol\chi_{x} \, \Sigma \cdot \e^5 \, \phi_x
  \nn\\
&& \phantom{spaci} + \sum\limits_{\mu=1}^4
 \left(\ol\chi_{x-\mu} \, \ol\Sigma \cdot \e^\mu \, \phi_{x}
 -\ol\phi_{x+\mu} \, \ol\Sigma \cdot \e^\mu \, \chi_x \right)
  +\ol\chi_{x} \, \ol\Sigma \cdot \e^5 \, \phi_x
-\ol\phi_{x} \, \ol\Sigma \cdot \e^5 \, \chi_{x} 
\Bigg],
\eea with $\Sigma=(\vec{\sigma}, -1)$ and $\ol\Sigma=(\vec{\sigma}, 1)$ and
 \beq
\psi_p = \begin{pmatrix}
\phi_p\\
\chi_p
\end{pmatrix},
\quad
\text{and}
 \quad
\ol \psi_p  
= 
\left( \ol \phi_p , \ol \chi_p  \right).\nn
\eeq 
With this two-component form of the action, examining whether or not this action has this symmetry is most easily accomplished by looking at a two-dimensional projection of this action and using the graphene picture.  In order for any of the terms to display this $\mathbb{Z}_5$ permutation symmetry, the choice of $B=1/\sqrt{5}$ and $C=1$ is required.  The first line of Eq.~\eqref{eq:BC_twocomponent} is given by the left figure in Fig.~\ref{GB} and shows the desired $\mathbb{Z}_3$, nearest neighbor behavior in this two dimensional projection (generalizes to the $\mathbb{Z}_5$ permutation symmetry in four dimensions).  However, the next line leads to the right figure in Fig.~\ref{GB}, which violates $\mathbb{Z}_3$ with next-to-nearest neighbor interactions (thus, violates the desired $\mathbb{Z}_5$ permutation symmetry in four dimensions).  Therefore, this action does not have the minimal symmetry required to eliminate the relevant operators.

\section{Other Non-Orthogonal Actions}
With the motivation of finding an action with exact chiral symmetry and this minimal $\mathbb{Z}_5$ permutation symmetry needed to eliminate the relevant operators, Ref.~\cite{bedaque2} explored several non-orthogonal actions which will be reviewed here.
\subsection{Modified Bori\c{c}i-Creutz Action}
With $B=1/\sqrt{5}$ and $C=1$, the first line of Eq.~\eqref{eq:BC_twocomponent} obeys the $\mathbb{Z}_5$ symmetry and the second line breaks this symmetry. A logical extension is to simply delete this last line.  Therefore, using the same notation as Eq.~\eqref{eq:BC_twocomponent}, this action with $B=1/\sqrt{5}$ and $C=1$ is given by
\bea
S_{MBC} 
&=& 
\frac{1}{2}\sum\limits_x 
\Bigg[
\sum\limits_{\mu=1}^4\left(
\ol\phi_{x-\mu} \, \Sigma \cdot \e^\mu \, \chi_{x}
 -\ol\chi_{x+\mu} \, \Sigma \cdot \e^\mu \, \phi_x  \right)
 +\ol\phi_{x} \, \Sigma \cdot \e^5 \, \chi_{x} 
 -\ol\chi_{x} \, \Sigma \cdot \e^5 \, \phi_x\Bigg].
\eea
However, the analysis of this action about the pole $p_\mu \simeq 0$ yields the behavior $\ol \psi (i\vec{\gamma}\cdot \vec{k} + \gamma_5\gamma_4 k_4) \psi$.   This behavior, refered to as mutilated fermions \cite{celmaster,Celmaster:1983jq}, is not the desired Dirac structure.

\subsection{``Hyperdiamond'' Action}
A further modification to the modified Bori\c{c}i-Creutz action to ensure the correct Dirac structure in the continuum limit and preserve at least the minimal $\mathbb{Z}_5$ permutation symmetry led to the ``hyperdiamond'' action, which is given by
\bea\label{eq:Z5}
S &=&  \sum_x \Bigg[\sum\limits_{\mu=1}^4\left(  \,
 \ol\phi_{x-\mu} \, \sigma \cdot \e^\mu \, \chi_{x}
 -\ol\chi_{x+\mu} \, \ol\sigma \cdot \e^\mu \, \phi_x \right)
 +\ol\phi_{x} \, \sigma \cdot \e^5 \, \chi_{x} 
 -\ol\chi_{x} \, \ol\sigma \cdot \e^5 \, \phi_x\Bigg],
\eea 
where 
\bea
\label{eq:es}
\e^1 &=& \frac{1}{4} (\phantom{+}\sqrt{5},\phantom{+}\sqrt{5},\phantom{+}\sqrt{5},\ 1),\nn\quad
\e^2 = \frac{1}{4} (\phantom{+}\sqrt{5},-\sqrt{5},-\sqrt{5},\ 1),\nn\quad
\e^3 = \frac{1}{4} (-\sqrt{5},-\sqrt{5},\phantom{+}\sqrt{5},\ 1),\nn\\
\e^4 &=& \frac{1}{4} (-\sqrt{5},\phantom{+}\sqrt{5},-\sqrt{5},\ 1),\nn\quad
\e^5 = -  ( \phantom{++}0,\phantom{++}0,\phantom{++}0,\,\ 1) .
\eea 
and $\ol\sigma=(\vec{\sigma}, -i)$, $\sigma=(\vec{\sigma}, i)$.  The key difference between this action and the modified Bori\c{c}i-Creutz action is the imaginary fourth component of the sigma four-vectors.  This alteration allows for the action to correctly reproduce the Dirac equation about $p_\mu \simeq 0$ (which is clear in Eq.~\eqref{eq:hyper_mom}).  In addition, the vectors $\e^\alpha$ satisfy the desired behavior needed for  $\mathbb{Z}_5$ (or $A_5$ or $S_5$) permutation symmetry, $\sum_\alpha \e^\alpha = 0$ and 
\bea\label{eq:e_properties}
\e^\alpha \cdot \e^\beta &=& \left\{ \begin{matrix}
                               \phantom{--}1 ,\ \  {\rm for} \ \  \alpha=\beta\\
                             -1/4,\ \  {\rm for} \ \  \alpha\neq\beta
                        \end{matrix} \right.
.\eea 
While the exact details of the transformation won't be covered here, one can show that this action has an $A_5$ permutation symmetry\footnote{See Ref.~\cite{bedaque2} for details.} invariant under pairs of permutations, which is enough symmetry to prevent the relevant operators from being generated.  In momentum space, the action is given by 
\beq \label{eq:hyper_mom}
S=
\int_p 
\ol\psi_p \left[ 
	i\sum_{\mu=1}^4 \sin(p_\mu) \e^\mu \cdot \gamma
	-\Big(\sum_{\mu=1}^4 \cos(p_\mu) \e^\mu+\e^5\Big) \cdot \gamma\, \gamma_5
\right] \psi_p
,\eeq 
with
\beq
\psi_p = \begin{pmatrix}
\phi_p\\
\chi_p
\end{pmatrix},
\quad 
\ol \psi_p  
= 
\left( \ol \phi_p , \ol \chi_p  \right),
\quad 
\text{and}
\quad 
\gamma_\mu
=
\begin{pmatrix}
0 & \sigma_\mu\\
 \ol\sigma_\mu  & 0 
 \end{pmatrix}
.\nn
\eeq 
From this form of the action, it is clear that it maintains exact chiral symmetry and correctly reproduces the Dirac equation in the continuum limit.  Additionally, due to the $A_5$ permutation symmetry, relevant operators will not be generated.  Unfortunately, this action yields excessive fermion doubling, not unlike the na\"ive fermion action.  An example of six poles in addition to the one at $p_\mu = 0$ is  $p_1 = -p_2 = -p_3 = p_4 = \cos^{-1}(-2/3)$. 

\section{Conclusion}
Non-orthogonal lattice actions can be used to enforce desirable features from a lattice action.  As shown for the Bori\c{c}i-Creutz action, clever discretizations can enforce exact chiral symmetry and minimal doubling.  However, upon gaining these benefits, additional symmetries are broken, which can (and often will) lead to the generation of relevant or marginal operators from radiative corrections.  An intricate balance of symmetry is needed in a discrete lattice action in order to have exact chiral symmetry, minimal doubling, and no relevant operators.  At this time, no non-orthogonal lattice action has accomplished this task.  However, this is by no means a proof that such a lattice action does not exist.  The possibility still stands that an action with just enough symmetry to rule out these relevant operators exists and if found, it would be a very efficient, cheap way to simulate chiral fermions.


  
\end{document}